\documentclass[11pt]{article}
\usepackage[latin1]{inputenc}
\usepackage[english]{babel}
\usepackage[namelimits]{amsmath}
\usepackage{authblk}
\usepackage{amssymb}
\usepackage{amsmath}
\usepackage{amsthm}
\usepackage{graphicx}

{\tiny}

\begin{document}
\title{{\bf A new class of anisotropic rotating fluids and some throat-like sources for Kerr metric
as examples}}
\author[1,2,3]{S. Viaggiu\thanks{s.viaggiu@unimarconi.it and viaggiu@axp.mat.uniroma2.it}}
\affil[1]{Dipartimento di Scienze Ingegneristiche, Universit\'a degli Studi Guglielmo Marconi, Via Plinio 44, I-00193 Roma, Italy.}
\affil[2]{INFN, Sezione di Roma 3, I-00146 Rome Italy.}
\affil[3]{Dipartimento di Matematica, Universit\`a di Roma ``Tor Vergata'', Via della Ricerca Scientifica, 1, I-00133 Roma, Italy.}
\date{\today}\maketitle

\begin{abstract}

Motivated by the increasing interest in finding physically viable rotating sources, we present a new 
class of anisotropic rotating solutions. The energy-momentum tensor compatible with the metric is composed of anisotropic matter with a non-vanishing 
energy flow around the symmetry axis and vanishing viscosity. 
The new class of solutions can be used to find new possible sources for the Kerr metric, to obtain new regular black hole solutions and to study galaxies with a central rotating black hole and an halo of dark matter. As an example, we obtain a 5-parameter class of solutions
representing
a two-way traversable wormhole smoothly matched to the Kerr one and satisfying all 
energy conditions outside the wormhole for a wide range of parameters, in particular for compact objects.
Finally, with a simple modification of the aforementioned solution, we obtain a source for Kerr
metric with a throat geometry, non-representing a two-way traversable wormhole and satisfying all energy conditions.   
\end{abstract}
{\it Keywords;} rotating fluid; anisotropic fluids; Kerr metric sources; wormholes;
energy conditions.

\section{Introduction}

A longstanding issue in general relativity is the one related to the search of 
exact metrics describing rotating, axially symmetric, isolated bodies. Since many celestial bodies in the universe are rotating, the argument is also of great interest in astrophysical context.
From a mathematical point of view, many methods have been developed in the 
literature (see for example \cite{1}-\cite{27} to cite someone) to build physically viable rotating interior metrics. However, in the literature, despite the efforts, only very few solutions are present
depicting global physically reasonable sources suitable for
isolated rotating bodies with astrophysical applications.\\ 
Since of the pioneering work in \cite{r1}, anisotropic sources have attracted a lot of interest.
First of all, anisotropies emerge for compact astrophysical objects as neutron stars and can lead 
to modifications of physical parameters  of the fluid composing the source 
\cite{r2,r3}. In the case of hypothesized boson stars, anisotropies \cite{r4}
naturally come into action. A long list of physical effects that can lead to anisotropies
can be found in \cite{r5}. In paticular, in \cite{r5} it has been argued
that, at least in the limit of slow rotation, the equation of hydrostatic equilibrium for a rotating source is given by the one of
an anisotropic fluid. There, anisotropies are proportional to the square of the angular velocity.
More recently, in \cite{r6} it has been shown that, both for spherical and axially symmetric 
astrophysical sources, stellar evolution processes, also for initial isotropic conditions, 
will always tend to form anisotropic fluid configurations. The results in \cite{r6} support the
idea that anisotropic effects during the dynamical evolution before equilibrium phase can still
remain in the final equilibrium configuration. As a consequence, anisotropies for compact objects
does not represent an anomaly, but rather the normality.\\
In \cite{24},
in order to obtain a physically viable anisotropic interior source for the Kerr metric,
a procedure suitable for the static case has been proposed, and thus extended to the stationary rotating one \cite{27}. The solution so obtained in \cite{27} has interesting properties.
In fact, the solution in \cite{27} is equipped with a non-vanishing energy-momentum flux in the equatorial plane.\\
An important issue concerning fluids configurations is given by energy conditions. 
However, to the best of my knowledge,
no regular rotating anisotropic sources for the Kerr metric representing astrophysical objects are known satisfying all energy conditions. It is thus evident the necessity to find new class of solutions for rotating sources.\\ 
More recently, the attention is also shifted, in order to find regular black hole solutions 
(see for example \cite{28}-\cite{32}) or to depict galaxies in terms of a central rotating black hole
sourronded by a dark matter halo, (see for example \cite{33}-\cite{36}) 
to the question to depict shadow effects due to black hole horizon. Papers in \cite{28}-\cite{36},
in order to obtain stationary solutions,
start from a static solution and then apply the Newman-Janis algorithm \cite{4}. Moreover, in \cite{37} has been found the rotating extension of traversable whormholes \cite{38} present in \cite{39}. Generally these solutions violate the week energy condition (WEC). In \cite{40} a solution with a traversable wormhole within the Einstein-Dirac-Maxell theory
is obtained, equipped with a non-phantom energy and a non-smooth gravitational field and matter on the throat with unusual properties also for fermions and shell mass. Solution in \cite{40} has been amended in \cite{41}, by means of an asymmetric spacetime with respect to the throat. However, it is not clear if such
a configuration depicted in \cite{41} can exist in the real word.\\
In this paper we generalize the metrics used, for example, in \cite{37}-\cite{41}. As an 
intriguing application, we obtain a 5-parameter solution smoothly matched to the Kerr one and 
representing a two-way traversable wormhole with matter content satisfying all energy conditions, week, strong and dominant
(WEC,SEC,DEC) for a wide range of parameters, in particular for compact astrophysical objects outside 
the wormhole. With a suitable modification, our solution can be interpreted as a source for Kerr 
metric with a throat geometry instead of the usual ring singularity of Kerr metric.
In section 2 the fundamental mathematical preliminaries are explained, while in section 3 the expression of the energy-momentum tensor $T_{\mu\nu}$ compatible with our solution is presented. In section 4 we present our new solution, while
in section 5
energy conditions for the aforementioned $T_{\mu\nu}$ are written and studied. In section 6 a possible source for Kerr without a traversable wormhole is presented and
studied.
Finally, in section 7 we outline some conclusions and final remarks.

\section{Mathematical preliminaries: transformations for the Kerr metric}

The Kerr metric in the Boyer-Lindquist coordinates $(t,r,x,\phi)$ with 
$\rho=\sqrt{r^2+a^2}\sqrt{1-x^2}, z=rx$ with $x=\cos\theta$ is given by:
\begin{eqnarray}
& & ds^2=-\left(1-\frac{2Mr}{\Sigma}\right)dt^2+\Sigma\left(\frac{dx^2}{1-x^2}
+\frac{dr^2}{\Delta}\right)-\frac{4Mra(1-x^2)}{\Sigma}dt d\phi+\nonumber\\
& &+\left(r^2+a^2+\frac{2Mra^2}{\Sigma}(1-x^2)\right)(1-x^2)d\phi^2, \label{1}\\
& & \Sigma=r^2+a^2x^2,\;\;\Delta=r^2-2Mr+a^2. \nonumber
\end{eqnarray}
As usual, the parameter $M$ denotes the ADM mass and $a$ the angular momentum for unit mass. Horizons are located at the roots of $\Delta=0$, while the ring singularity is given by the zero of $\Sigma$, 
i.e. $r=0, x=0$.\\
In \cite{29}, in order to study regular black holes, the metric (\ref{1}) is used with
$M\rightarrow m(r)$. In \cite{10} it has been shown that with such a transformation, the (\ref{1})
does not describe a vacuum meric but rather an anisotropic rotating source equipped with an energy density $E$, a radial pressure $P_r$ and the tangential pressures $P_x$ and $P_{\phi}$. In \cite{28}
is also considered, within a generalized Newman-Janis algorithm, the case where $a\rightarrow a(r)$. It can be shown that the energy momentum so generated is equipped with anisotropic matter, heat flow and also a non vanishing viscosity term. The metric so generated in \cite{28} fails to describe sources 
satisfying al least the WEC.\\
In this paper we explore also the transformation $r\rightarrow f(r)$. The final form for the metric 
becomes:
\begin{eqnarray}
& & ds^2=-\left(1-\frac{2m(r)f(r)}{\Sigma}\right)dt^2+\Sigma\left(\frac{dx^2}{1-x^2}
+\frac{dr^2}{\Delta}\right)-\frac{4m(r)f(r)a(1-x^2)}{\Sigma}dt d\phi+\nonumber\\
& &+\left({f^2(r)}+a^2+\frac{2m(r)f(r)a^2}{\Sigma}(1-x^2)\right)(1-x^2)d\phi^2, \label{2}\\
& & \Sigma=f^2(r)+a^2x^2,\;\;\Delta=f^2(r)-2m(r)f(r)+a^2. \label{3}
\end{eqnarray}
To the best of my knowledge the generalization (\ref{2}) has not been considered in literature. In particular, metrics in \cite{37}-\cite{41} are special case of (\ref{2}). As an example, the metric in
\cite{37} can be obtained from (\ref{2}) with $f(r)=\sqrt{r^2+{\ell}^2},\;\ell\in\Re$
together with $m(r)=M$.\\
First of all, we check the "elementary flatness" condition on the rotation axis at 
$x=1$:
\begin{equation}
\lim_{x\rightarrow 1} \frac{{g}_{\phi\phi,\mu}\;{{g}_{\phi\phi}}^{,\mu}}{4\;{g}_{\phi\phi}}=1,
\label{4}
\end{equation}
with ${g}_{\phi\phi}$ given by (\ref{2}). The regularity condition (\ref{4}) holds for any 
function $f(r)$, provided that $f(r)$ is a bounded function near the axis at $r=0$. The last condition
on $f(r)$ assures that also for $r\rightarrow 0$, condition (\ref{4}) is fulfilled.\\
If we are searching for solutions that are regular, without ring singularities, horizons and infinite redshift surfaces characterizing the Kerr metric, it is sufficient (see section 6
for more details) to satisfy the following conditions:
\begin{equation}
f(r)>2M,\;\;\;m(r)\leq M\;\;\forall r\in[0,+\infty).
\label{5}
\end{equation}
If we are interested to obtain solutions smoothly matched to the Kerr one at some finite radius
$R$, then the (\ref{5}) is still valid but $\forall\in[0,R]$. With only the condition 
$f(r)>0$ we obtain regular black holes. Moreover, one must impose that, for $m(r)=0$,  
$f(r)$ reduces to $f(r)=r$. This is because in this case the flat metric in spheroidal coordinates 
must be regained. Finally, in order to smoothly match the (\ref{2}) at $r=R$, we must have:
\begin{equation}
f(R)=R,\;f_{,r}(R)=1,\;m(R)=M,\;m_{,r}(R)=0.
\label{5a}
\end{equation}
Obviously, we can also obtain interior asymptotically flat anisotropic solutions without imposing the
boundary conditions (\ref{5a}) but instead:
\begin{equation}
\lim_{r\rightarrow\infty}\frac{f(r)}{r}=1,\;\;\;\lim_{r\rightarrow\infty}
m(r)=M,
\label{r3}
\end{equation}
where $M$ is the ADM mass of the spacetime.\\
Summarizing, with metric (\ref{2})-(\ref{3}) it is possible, by adopting
conditions (\ref{5}), to obtaing regular sources matter field that can be matched, qith (\ref{5a}),
to tha vacuum Kerr metric. It is also possibkle to study regular black holes solutions, but our main interest in this paper is devoted to rotating nopn black holes configurations.\\
We are ready to specialize the energy momentum tensor compatible with metric (\ref{2}).

\section{Energy-momentum tensor compatible with the chosen metric}

It is known \cite{29} that the (\ref{2}) with $f(r)=r$ is provided by anisotropic matter 
$(E,P_r,P_x,P_{\phi})$. First of all, we note that the non-vanishing components of 
$G_{\mu\nu}=-T_{\mu\nu}$ are $G_{tt}, G_{t\phi}, G_{rr}, G_{xx}, G_{\phi\phi}$. Since
$G_{rx}=0$, as a consequence we have certainly a vanishing viscosity.
In order to study the $T_{\mu\nu}$ compatible with (\ref{2}), we 
introduce the orthonormal tetrad 
$\{V_{(t)}^{\mu}, S_{(r)}^{\mu}, L_{(x)}^{\mu}, W_{(\phi)}^{\mu} \}$ with
\begin{equation}
g_{\mu\nu}=-V_{(t)\mu}V_{(t)\nu}+S_{(r)\mu}S_{(r)\nu}+L_{(x)\mu}L_{(x)\nu}+
W_{(\phi)\mu}W_{(\phi)\nu},
\label{6}
\end{equation}
and
\begin{eqnarray}
& & V_{(t)}^{\mu}=
\left[\frac{{f(r)}^2+a^2}{\sqrt{\Sigma\Delta}}, 0, 0, \frac{a}{\sqrt{\Sigma\Delta}}\right],
\label{7}\\
& & S_{(r)}^{\mu}=
\left[0, \sqrt{\frac{\Delta}{\Sigma}}, 0, 0\right],\nonumber\\
& & L_{(x)}^{\mu}=
\left[0, 0, \sqrt{\frac{1-x^2}{\Sigma}}, 0\right],\nonumber\\
& & W_{(\phi)}^{\mu}=
\left[\frac{a(1-x^2)}{\sqrt{\Sigma(1-x^2)}}, 0, 0, \frac{1}{\sqrt{\Sigma(1-x^2)}}\right],
\end{eqnarray}
with $\Sigma$ and $\Delta$ given by (\ref{3}). Thus we have:
\begin{equation}
T_{\mu\nu}=EV_{(t)\mu}V_{(t)\nu}+P_rS_{(r)\mu}S_{(r)\nu}+P_xL_{(x)\mu}L_{(x)\nu}+
P_{\phi}W_{(\phi)\mu}W_{(\phi)\nu}.
\label{8}
\end{equation}
It is easy to verify that metric (\ref{2}) is compatible with (\ref{8}) if and only if
\begin{equation}
f(r) {f(r)}_{,r,r}+{f(r)}_{,r}^{2}-1=0. 
\label{9}
\end{equation}
With $f(r)$ satisfying the (\ref{9}) we obtain a vanishing heat flow term $K$. With the simplest solution
of (\ref{9}) $f(r)=r$, we obtain the metric in \cite{29}. Another solution, merely representing
a coordinate transformation, is $f(r)=r+const$. Another possible solution is 
$f(r)=\sqrt{r^2+\ell^2}$, thus obtaining the metric in \cite{37}. Hence, we conclude that the 
$T_{\mu\nu}$ compatible with (\ref{2}) is:
\begin{eqnarray}
& & T_{\mu\nu}=EV_{(t)\mu}V_{(t)\nu}+P_rS_{(r)\mu}S_{(r)\nu}+P_xL_{(x)\mu}L_{(x)\nu}+
P_{\phi}W_{(\phi)\mu}W_{(\phi)\nu}+\nonumber\\
& &+ K\left(V_{(t)\mu}W_{(\phi)\nu}+V_{(t)\nu}W_{(\phi)\mu}\right).
\label{10}
\end{eqnarray}
The last term in (\ref{10}) describes 
a non-vanishing 
energy flow around the symmetry axis.
With (\ref{10}), after 
denoting with $V_{(t)t}=A, V_{(t)\phi}=B, W_{(\phi)t}=C, W_{(\phi)\phi}=D$,
the Einstein's field equations become (the expressions for $G_{\mu\nu}$ are not explicitely written
but have been checked with Maple):
\begin{eqnarray}
& & A^2 E+C^2 P_{\phi}+2AC K=-G_{tt}, \label{11}\\
& & AB E+CD P_{\phi}+K(AD+BC)=-G_{t\phi},\label{12}\\
& & B^2 E+D^2 P_{\phi}+2BD K=-G_{\phi\phi}, \label{13}\\
& & P_{r}=-\frac{\Delta}{\Sigma} G_{rr}, \label{14}\\
& & P_{x}=-\frac{(1-x^2)}{\Sigma}G_{xx}. \label{15}
\end{eqnarray}
From equations (\ref{11})-(\ref{13}), we can solve for $E, P_{\phi}, K$ as a linear system with 
coefficients $A,B,C,D$. Cramer's rule can be applied with a unique solution for $E, P_{\phi}, K$.
Concerning $K$ we obtain:
\begin{equation}
K=\frac{a\sqrt{\Delta(1-x^2)}\left[f f_{,r,r}-1+f_{,r}^2\right]}{\Sigma^2}.
\label{k}
\end{equation}
We obtain $K=0$ provided that (\ref{9}) is satisfied.
The presence of 
a non-vanishing 
energy flow term around the symmetry axis
given by (\ref{k}) 
is an interesting general feature of our model represented by the metric (\ref{2}) with
(\ref{3}). To this pourpose, it should be noticed that also in the solution presented in \cite{27}
a non-vanishing 
energy flow around the symmetry axis is present. The physical origin of the term (\ref{k}) is very similar to the one present in \cite{27}. 
From (\ref{k}) it follows that $K$ is vanishing for $a=0$, i.e. it comes in action only for rotating sources and consequently in the static limit such a term is absent. 
This flow
is also vanishing for $x=\pm 1$, i.e. on the rotation axis at $\rho=0$ and for any 
surface at $r=constant$ it reaches 
the maximum value at the equatorial plane at $x=0$, i.e. $z=0$. Moreover, if we consider regular black hole solutions, energy flow is vanishing at the horizons located at $\Delta=0$. As noted also
in \cite{27}, the presence of a non-vanishing $K$ is a consequence of the rotation of the source and an analogy with stationary Einstein-Maxwell systems \cite{42,43}, where a non-vanishing component of Poyinting vector of 
electromagnetic nature arises, can be suggested.\\
Moreover, with (\ref{2}), we generally obtain solutions with:
\begin{equation}
P_{x}-P_{\phi}=\frac{2a^2f^2(1-x^2)(f_{,r}^2-1)}{\Sigma^3}.
\label{k2}
\end{equation}
In the next section we present a new rotating source with a traversable wormhole.

\section{Generating wormhole solutions}
As an intriguing application, we can consider regular sources for the Kerr metric representing 
two-way
trasversable rotating wormholes \'{a} la Morris-Thoarne \cite{38}, thus of the kind discussed in \cite{37} and \cite{39}.
In the range $r\in[-R,R]$ we can consider:
\begin{eqnarray}
& & f(r)=|r|+kM-\frac{cM}{R^2}{(R-|r|)}^2,\;k\geq 0,\;c\geq 0,\;\;R > 0,\label{23}\\
& & m(r)=M-\frac{M}{R^2}{(R-|r|)}^2.\nonumber
\end{eqnarray}
First of all, note that solution (\ref{23}) is symmetrical and 
smoothly matched to the Kerr one in both
"universes" at $|r|=R$. At $|r|=R$ we can perform the coordinate transformation 
$|r|+kM=\overline{r}$ and recast Kerr metric in the usual form (\ref{1})  with
$r\rightarrow\overline{r}$.
Note that solution (\ref{23}) is continuos at $r=0$ (spacetime is connected) but not differentiable.
Nevertheless, the following equations hold:
\begin{eqnarray}
& & E(r\rightarrow 0^+)=E(r\rightarrow 0^-), \label{25}\\
& & K(r\rightarrow 0^+)=K(r\rightarrow 0^-), \nonumber\\
& & P_{\phi}(r\rightarrow 0^+)=P_{\phi}(r\rightarrow 0^-), \nonumber\\
& & P_r(r\rightarrow 0^+)=P_r(r\rightarrow 0^-), \nonumber\\
& & P_x(r\rightarrow 0^+)=P_x(r\rightarrow 0^-), \nonumber
\end{eqnarray}
Conditions (\ref{25}) does imply that 
$T_{\mu\nu}(r\rightarrow 0^+)=T_{\mu\nu}(r\rightarrow 0^-)$. The (\ref{25}) however does not
imply that distributional source is not present at $r=0$, as we will see leter. 
Conversely, the 
quantities $E_{,r}, P_{\phi,r}, K_{,r}, P_{x,r}, P_{r,r}$ are discontinuous, but not the modulus.
In practice:
\begin{eqnarray}
& & E_{,r}(r\rightarrow 0^+)=-E_{,r}(r\rightarrow 0^-), \label{26}\\
& & K_{,r}(r\rightarrow 0^+)=-K_{,r}(r\rightarrow 0^-), \nonumber\\
& & P_{\phi,r}(r\rightarrow 0^+)=-P_{\phi,r}(r\rightarrow 0^-), \nonumber\\
& & P_{r,r}(r\rightarrow 0^+)=-P_{r,r}(r\rightarrow 0^-), \nonumber\\
& & P_{x,r}(r\rightarrow 0^+)=-P_{x,r}(r\rightarrow 0^-), \nonumber
\end{eqnarray}
Equations (\ref{26}) does imply that modulus of the velocity of modes is conserved 
by crossing the throat.\\
Solution so generated has 5 free parameters $\{R, M,a,k,c\}$.
In the next section we study regularity and energy conditions.

\section{Regularity and Energy conditions}
First of all we study regularity of (\ref{23}). If we consider solutions with no
singularities and without horizons within, we must impose the condition $f(r)>2M\; \forall r\leq R$.
The last condition is certainly satisfied (sufficient condition) provided that
(for more details see section 6) 
\begin{equation}
k-c>2.
\label{24} 
\end{equation}
For the choice of $\{k,c\}$ such that $k-c>0$, we obtain regular black holes solutions.\\
Since in this paper we are mainly interested to obtain
physically viable regular rotating sources for (\ref{1}) without horizons, we limitate to cases
with condition (\ref{24}).\\
In the following we analyse energy conditions.
Since solution (\ref{23}) is symmetrical with respect to $r$, it is sufficient to study energy 
conditions outside the throat in the range $r>0$.
As a first necessary step, we study energy conditions suitable for the metric 
(\ref{2}). The eigenvalues equation to solve is $|T_{\mu\nu}-\lambda g_{\mu\nu}|=0$. Energy conditions \cite{44,45} are exptected to hold for all ordinary matter present in the universe and consequently also for a physically viable ordinary rotating source. Generally, it is not a simple task to satisfy 
all energy conditions, in particular for rotating regular anisotropic sources. To the best of my knowdlege, no examples of
regular rotating anisotropic sources satisfying all energy conditions are present in the literature. 
With (\ref{7}) and (\ref{10}) we obtain:
\begin{eqnarray}
& & \lambda_{t}=
\frac{1}{2}\left(P_{\phi}-E-\sqrt{{(E+P_{\phi})}^2-4K^2}\right),\label{16}\\
& & \lambda_{\phi}=
\frac{1}{2}\left(P_{\phi}-E+\sqrt{{(E+P_{\phi})}^2-4K^2}\right),\label{17}\\
& & \lambda_{r}=P_{r}, \label{18}\\
& & \lambda_{x}=P_x. \label{19}
\end{eqnarray}
As a consequence, according to \cite{44,45,23}, for the WEC we have:
\begin{equation}
-\lambda_{t}\geq 0,\;\;\;-\lambda_{t}+\lambda_{i}\geq 0,
\label{20}
\end{equation}
for the SEC:
\begin{equation}
-\lambda_{t}+\sum_{i}\lambda_{i}\geq 0,\;\;\;-\lambda_{t}+\lambda_{i}\geq 0,
\label{21}
\end{equation}
while for DEC:
\begin{equation}
-\lambda_{t}\geq 0,\;\;\;\lambda_{t}\leq \lambda_{i}\leq -\lambda_{t}.
\label{22}
\end{equation}
WEC does imply positivity of the energy for any observer. SEC says that pressures cannot have too large 
values, while DEC does imply the WEC with the request that modes propagate at velocity less than the one 
of the light in vacuum. For a viable solution representing realistic sources with usual matter content, all conditions (\ref{20})-(\ref{22}) should be satisfied.\\ 
Since our model contains 5 parameters, a complete study of energy conditions is not a simple task. However, some general arguments can be done in order to show that for a wide range of 
parameters energy conditions can be fulfilled for $r\neq 0$.\\
More generally, from a preliminary study, it seems that a necessary condition to satisfy energy conditions is to provide monotonically increasing expressions for $f(r)$ and $m(r)$. Although we failed to show this proposition, this looks like a reasonable statement. This seems natural for mass 
function $m(r)$, while $f(r)$ in our metric (\ref{2}) "substitutes" the radial coordinate 
$r$ in (\ref{1}). Moreover, condition $f_{,r,r}(r)<0$ is required \cite{46} as a necessary one
for energy conditions.
Solution (\ref{23}) satisfies these preliminary conditions. We thus are ready to study energy conditions
for $r>0$ that in turn, thanks to symmetry of (\ref{21}), is also equivalent to the analogue study for $r<0$.\\
To start with, we study the energy density $E$. To analyse positivity of $E$, we must firstly consider 
equation $E(r\rightarrow 0^{+},x)>0$. The aforementioned equation could impose restrictions on a possible maximum allowed value for $R$, i.e. $R_{max}$. From a preliminary study, we have that no limitation arises 
for $c<<1$. Limitations can arise for $c\simeq 1$ it depending on the value of 
$k$. As an example,
for $c=0.9, k=3$ and $a=\frac{1}{2}$ we have $R_{max}\simeq 1385 M$. The role of $a$, in this case is to 
decrease such value. For example, for $a=1,c=0.9, k=3$ we have $R_{max}\simeq 8.8 M$. 
Moreover, for $a=1/2, k=10, c=3/2$ we have $R_{max}\simeq 26.5M$, while for
$a=1, k=10, c=3/2$ we obtain $R_{max}\simeq 24.9 M$. These estimations show that the solution 
(\ref{27}) is particularly suitable to depict very compact objects as neutron stars.\\
In figure 1 we plotted 
$E$ for $a=1/2, M=1, R=10, c=3/2, k=10$.
\begin{figure}
\centering
\includegraphics[scale=0.65]{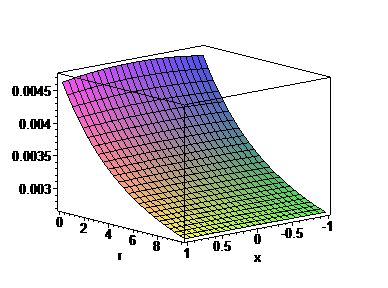} \quad
\caption{Plot of $E$ for $a=1/2, M=1, c=3/2, k=10, R=10$} 
\label{F1}
\end{figure}
Concerning the radial pressure $P_r$, as dictated by junction conditions at $r=R$,
we always have $P_{r}(r=R)=0$, while for $r\rightarrow 0^{+}$ we obtain:
\begin{equation}
P_r(r\rightarrow 0^{+})=
\frac{4M^3{(k-c)}^2(Rc+Mc^2-2Mc-R)}{R^2{(c^2M^2+k^2M^2-2kM^2c+a^2 x^2)}^2}.
\label{28}
\end{equation}
Regularity of (\ref{28}) requires that $k>2$. Since, in order to avoid black hole solutions, 
we have $R>2M$, as a result of (\ref{28}) we have that $P_r>0$ for $c>1$ and $P_r<0$
for $c\leq 1$. The behavior of $P_r$ for $c>1$ is very similar to the one depicted in figure 1 and with opposite sign for $c<1$. Concerning $P_x$ and $P_{\phi}$, they have opposite sign with respect to 
$P_r$: when $P_r<0, \{P_x, P_{\phi}\}>0$  and for $P_r>0, \{P_x, P_{\phi}\}<0$.\\
As stated above, a complete study for the 5-parameters model considered is not a simple task. 
However, a preliminary study shows that all energy conditions can be fulfilled for a wide range of the
parameters, in particular for compact objects with $R/M<100$. In the following we analyse in particular
the case with $c>1$. This is because in this way radial pressure $P_r$ is positive
inside the source.
The positivity of $P_r$ is expected for usual physically viable sources. Moreover, the positivity
of $P_r$ is required to enhance the stability of the configuration. With this choice we have 
$\{P_{\phi}, P_x\}<0$. The strategy to fulfill energy conditions is the following.
For WEC (\ref{20}), once established the positivity of $E$, the negativity
of eigenvalue (\ref{16}), i.e. $-\lambda_{t}>0$ must be verified. After that, since 
$P_r>0$, one must check the positivity of $-\lambda_{t}+P_x$. The quantity 
$-\lambda_t+\lambda_{\phi}$ becomes, thanks to (\ref{16}) and (\ref{17}),  
$\sqrt{{(E+P_{\phi})}^2-4K^2}$ that is always positive in its domain. For the SEC
(\ref{21}),  one must only check the condition 
\begin{equation}
\sqrt{{(E+P_{\phi})}^2-4K^2}+P_r+P_x\geq 0.
\label{Ec1}
\end{equation}
Finally, concerning the DEC (\ref{22}), since $-\lambda_{t}>0$ for the WEC and in our case 
$P_r>0$, it is sufficent to verify that
\begin{equation}
\frac{|\lambda_{\phi}|}{-\lambda_{t}}<1,\;
\frac{|P_{x}|}{-\lambda_{t}}<1,\;
\frac{P_r}{-\lambda_{t}}<1.
\label{Ec2}
\end{equation}
By following the aforementioned strategy to study energy conditions, we present the case 
with $M=1, R=10, c=3/2, k=10, a=1/2$ in figures 2,3,4,5,6,7. In figures 2,3,4 we test WEC and SEC.
Note that, since quantities involved are small, to highlights positivity conditions required for WEC and SEC we raise quantities  to a power of $1/20$. Figures 5,6,7 refer to DEC, while in figures 8,9 we plotted 
$|P_{x}|/(-\lambda_t)$ respectively for $a=1$ and $a=0$. 
Finally, in figures 
8,9 we show that results do not change in a significative way also for low and hight values of $a$.
To this regard, similar plot to 8,9 follow for the other quantities involved in WEC, SEC and DEC.\\
The next step is the study of energy conditions at $r=0$. 
Solution (\ref{23}) has some distributional source at 
$r=0$. In fact, at $r=0$ thanks to (\ref{25}), we find a deltaform expression for $T_{\mu\nu}$,
i.e. $T_{\mu\nu}(r=0)\sim\delta(r)$, with $\frac{d\Theta}{dr}=\delta(r)$, 
$H(r)=2\Theta(r)-1$ the Heaviside function and
$\Theta(r)$ is the step function. Hence, the delta distribution for 
$T_{\mu\nu}(r=0)$ arises from second partial derivative of $f(r)$ and $m(r)$ calculated at 
$r=0$. After rembering that $r\delta(r)=0$ and denoting  $f(0)=M(k-c)=s$ 
and $2+\frac{4cM}{R}=Z $ we get that at $r=0$ 
$P_r=0, P_x=P_{\phi}$ together with:
\begin{eqnarray}
& & E=\frac{1}{{(s^2+a^2 x^2)}^3}
\left[-2s^5-2s^3x^2 a^2-2s^3a^2-2sx^2 a^4\right]Z\delta(r),
\nonumber\\
& & P_{\phi}=\frac{\left[\left(a^4 x^4 s+2s^3 x^2 a^2+s^5\right)Z-\left(
4 a^4 x^4 s+8a^2 x^2 s^3+
4s^5\right)\frac{M}{R}\right]}{{(s^2+a^2 x^2)}^3}
\delta(r),\nonumber\\
& & K=\frac{Zas\sqrt{1-x^2}\sqrt{s^2+a^2}}{{(s^2+a^2 x^2)}^2}
\delta(r)\label{6.1}.
\end{eqnarray} 
From inspection of (\ref{6.1}), in particular for $E$, it is easy to see that a necessary condition
to satisfy energy conditions is that $Z\leq 0$, i.e. $R\leq -2cM$. Hence energy conditions require that 
$c<0$. It is easy to show that it is not possible to find solutions with
$c<0$ and $k-c>2$ that can at the same time satisfy energy conditions both for $r\neq 0$
and $r=0$. Nevertheless, can be of interest the study of energy conditions 
of the wormhole located at $r=0$. The most interesting case is the one with $Z=0$.
From (\ref{6.1}) we have $E=K=0$ together with
$P_{\phi}=P_x<0$ and $-\lambda_1=-P_{\phi}>0$, $\lambda_{\phi}=0$, $-\lambda_1+P_x=0$
and also $P_x/(-\lambda_1)=1$ and as a consequence all energy conditions are satisfied. In this case we have a null fluid with a vanishing eigenvalue ($\lambda_{\phi}$) and with modes propagating at the speed 
of light ($P_x/(-\lambda_1)=1$). This is typical of a radiation-like fluid.\\
Also for $R<-2cM$ we have a range of parameters such that at least WEC is satisfied for the wormhole
at $r=0$.\\
The Penrose diagram of (\ref{23}) is provided by two Minkowski spacetimes with a wormhole at $r=0$, 
in a similar way to the two way traversable wormhole in \cite{37}.\\
Summarizing, we obtained a wormhole solution where the violation of energy conditions is minimal and 
located at the throat only at $r=0$. This is an interesting feature of solution (\ref{23}). 

\begin{figure}
\centering
\includegraphics[scale=0.35]{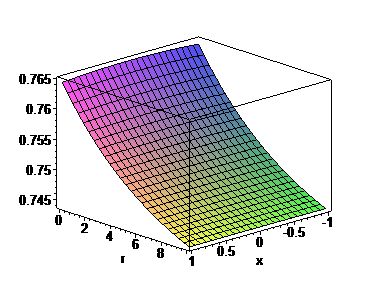} \quad
\caption{In figure the plot of ${(-\lambda_t)}^{1/20}$ for $M=1, R=10, c=3/2, k=10, a=1/2$.} 
\label{F2}
\includegraphics[scale=0.35]{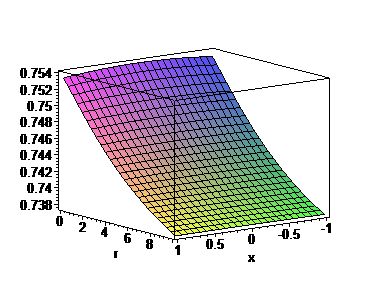}\\ 
\caption{In figure the plot of ${(-\lambda_t+P_x)}^{1/20}$ for $M=1, R=10, c=3/2, k=10, a=1/2$.}
\label{F3}
\includegraphics[scale=0.35]{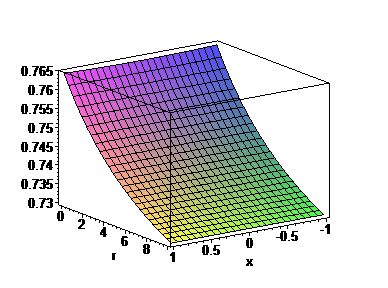}
\caption{In figure the plot of ${(-\lambda_t+\sum_i \lambda_i)}^{1/20}$ for $M=1, R=10, c=3/2, k=10, a=1/2$.}
\label{F4}
\includegraphics[scale=0.35]{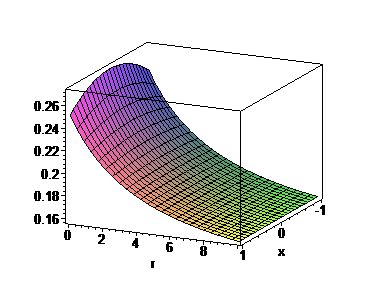}\\ 
\caption{In figure the plot $|\lambda_{\phi}|/(-\lambda_t)$ for $M=1, R=10, c=3/2, k=10, a=1/2$.}
\label{F5}
\end{figure}

\begin{figure}
\centering
\includegraphics[scale=0.35]{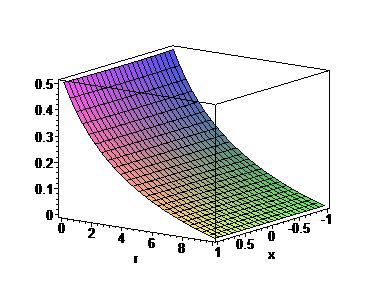} \quad
\caption{In figure the plot of $P_{r}/(-\lambda_t)$ for $M=1, R=10, c=3/2, k=10, a=1/2$.} 
\label{F6}
\includegraphics[scale=0.35]{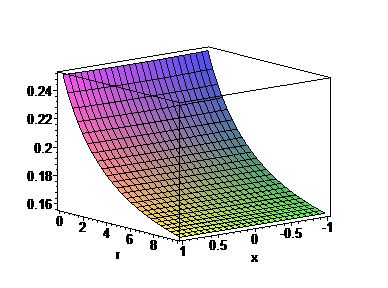}\\ 
\caption{In figure the plot of $|P_{x}|/(-\lambda_t)$ for $M=1, R=10, c=3/2, k=10, a=1/2$.}
\label{F7}
\includegraphics[scale=0.35]{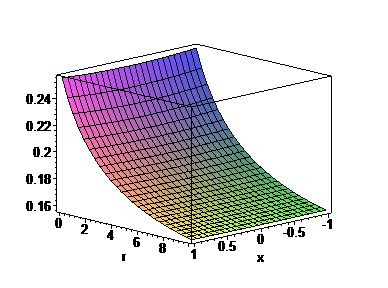}
\caption{In figure the plot of $|P_{x}|/(-\lambda_t)$ for $M=1, R=10, c=3/2, k=10, a=1$.}
\label{F8}
\includegraphics[scale=0.35]{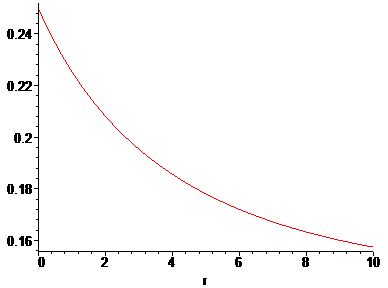}\\ 
\caption{In figure the plot $|P_{x}|/(-\lambda_t)$ for $M=1, R=10, c=3/2, k=10, a=0$.}
\label{F9}
\end{figure}

\section{A possible source for Kerr metric with a throat geometry at $r=0$}

As a further example, we consider the following solution with $r\in[0,R]$ obtained with a simple modification of (\ref{23});
\begin{eqnarray}
& & f(r)=r+kM-\frac{cM}{R^2}{(R-r)}^2,\;k\geq 0,\;c\geq 0,\;\;R > 0,\label{27}\\
& & m(r)=M-\frac{M}{R^2}{(R-r)}^2,\nonumber 
\end{eqnarray}
that in particular
can be used as a particular interior source for Kerr metric. In fact, the (\ref{2})
with (\ref{27}) can be smoothly matched to Kerr solution at $r=R$.\\
Also for (\ref{27}), if we consider solutions with no
singularities and without horizons and ergoregions within, we must impose the condition $f(r)>2M\; \forall r\leq R$ that is fullfilled with (\ref{24}). Concerning energy cobditions, obviously all consideration of section above are still valid and solution (\ref{27}) satisfies energy conditions
for a wide range of parameters, in particular for compact objects.
Hence, all plots of section 5 are obviously valid for (\ref{27}.)\\
As a next step, we analyse regularity of solution (\ref{27}). 
Preliminarily, note that 
$m(r=0)=0$ and $m(r)$ is monotonically increasing an differentiable for $r\in[0,R]$ and also $f(r)$ is monotonically incresing, with $c>0$, and differentiable for $r\in[0,R]$. Solution (\ref{27}),
thanks to the continuity of the first and second fundamental form at the boundary
$r=R$, is smoothly matched to the 
Kerr one at $r=R$\footnote{There, with coordinate transformation 
$\overline{r}=r+kM$ the usual expression for Kerr metric is regained for $r>R$.}. 
Hence, for $r\geq R$ the solution is the Kerr one with $r\geq R>2M$, i.e.
without the ring singularity, horizons and ergoregions.\\
To be more quantitative, note that conditions (\ref{5a}) garantee the continuity of the first and second
fundamental form at $r=R$. It is easy to see that solution (\ref{27}) satisfies (\ref{5a}).\\ 
Concerning regularity, the ring singularity of the Kerr solution is located at $\Sigma=0$. With respect to
(\ref{2}), (\ref{3}), singularity is present for $\Sigma=f^2(r)+a^2x^2=0$. With solution 
(\ref{27}), since  $m(r)\leq M\;\forall r\in[0,R]$, we have that a necessary and sufficient condition to avoid
the ring singularity of the Kerr solution is that $f(r)>0\;\forall r\in[0,R]$. With (\ref{27}) and 
thanks to monotony of $f(r)$ depicted above, we can
write a sufficient condition given by $k-c>0$. It is easy to show that with $k-c>0$
together with finiteness of $m(r), f(r)$ in (\ref{27}) and their first and second derivatives, 
no singularities are present and scalar invariants (Ricci, Kretschmann ...) are well defined 
$\forall r\in[0,R]$
\footnote{It is not necessary to write down explicitely the long expressions for scalars, but all look like $\sim\frac{1}{\Sigma^n}$, with $n$ some positive integer.}. 
Moreover, elementary flatness condition (\ref{4}) is obviously verified for our solution.\\
If we want to describe a regular source for the exterior Kerr metric
(\ref{1}) without a regular black hole within (and outside with
$R>2M$), the solution (\ref{27}) must 
be also free of horizons and ergoregions. Horizons are located at $\Delta=0$. 
Thanks to (\ref{3}), this happens when:
\begin{equation}
f(r)=m(r)\pm\sqrt{m^2(r)-a^2},\;\;r\in[0,R].
\label{r1}
\end{equation}
For the aforementioned properties of $f(r)$ and $m(r)$ given by (\ref{27}), a sufficient condition to avoid horizons is $f(r)>2M$, i.e. the (\ref{24}). The boundary of ergoregion, dubbed ergosurface,
is given by equation $g_{tt}=0$ that in our case becomes:
\begin{equation}
f^2(r)-2m(r)f(r)+a^2 x^2=0,\;\;r\in[0,R].
\label{r2}
\end{equation}
Ergoregions and infinite redshift surfaces are thus absent provided that 
$\Sigma-2m(r)f(r)>0\;\forall r\in[0,R]$.
A sufficient but not necessary condition to avoid ergoregions is thus again, after 
dropping the term $a^2x^2$ in $\Sigma$, given by $f(r)>2M$, i.e. the (\ref{24}).\\
Summarizing, with condition (\ref{24}) our source for Kerr metric is regular and free of horizons, 
ergoregions and infinite redshift surfaces. Moreover, we always 
have that $g_{\phi\phi}>0$ and no closed timelike curves are present.\\
A further important question is related to the structure of the solution at $r=0$. 
Note that since we are using oblate coordinates, the centre of solution (\ref{27}) is located on the
axis at $x=\pm 1$ with  $r=0$. Solution (\ref{27}) is the same as (\ref{23}) but without the region $r<0$ and the distributional source at $r=0$ and with a throat
at $r=0$.
As a consequence, for the study of section 5, the (\ref{27}) is a regular source for Kerr
metric satisfying all energy conditions for a wide range of parameters, in particular for compact objects
where anisotropic fluids are aspected to come in action. 
Moreover, for (\ref{27}) the difference $k-c$ can be interpreted as a measure of 
thickness of the throat. As usual, throat is located at the minimum of $f(r)$ that in our case is 
found at $r=0$.\\
From a mathematical point of view, our solution could be extended for negative 
values of $r$. However, in this case a curvature ring singularity, similar to the Kerr one, is present for $x=0$ ($z=0$) and at the negative root of $f(r)$. It is elementar to see that the negative root 
for $f(r)$ does happen for some negative $r^{*}$ with $r^{*}\in (-kM,0)$.\\
As a result, as far as the radial coordinate remains non negative, solution (\ref{27}) represents a 
regular source for Kerr metric with a throat at $r=0$. The aforementioned throat does not represent,
differently from solution (\ref{23}), a traversable wormhole.

\section{Conclusions and final remarks}

In this paper we have presented a new class of metrics that generalize, for example, the ones used in
\cite{37}-\cite{41}. The solution so generated can describe regular anisotropic rotating fluids with
vanishing viscosity. Such solutions can be used to obtain sources for Kerr metric, to study regular black holes, to build solutions, in order to mimics galactic halo, with rotating black holes sourronded by dark matter and also transversable wormholes.
Note that metric (\ref{2}) can also be generalized
by considering instead of the (\ref{1}) the Kerr-Newman solution, thus introducing an electric charge
$Q$. This can be obtained with the transformation 
$m(r)\rightarrow m(r)-\frac{Q^2}{2f(r)}$.\\
As an example for metric (\ref{2}), a 5-parameter solution, namely the (\ref{23}), has been presented.
The aforementioned solution describes
two symmetrical universes, for  $r<0$ and 
$r>0$, smoothly matched to the Kerr one in both universes. This solution 
represents a two-way traversable wormhole. Although the spacetime 
so obtained  contains a distributional matter at $r=0$ with generally phantom 
energy with respect to the solution at $r\neq 0$, energy-momentum tensor is continuous at $r=0$, and as a result matter fields 
match continuosly the two "universes". Moreover, energy conditions are fulfilled for $r\neq 0$, i.e.
outside the throat, in particular for compact objects where anisotropies are expected to come into action. This is an interesting property of our model since energy conditions are violated only for the
throat, i.e. in a minimal way.\\
With a suitable modification we obtain solution (\ref{27}).
Solution (\ref{27}) can be interpreted as a particular source of Kerr metric, satisfying all energy
conditions, with the ring singularity of Kerr metric substituted with a throat that does not 
represent a two-way traversable wormhole.
Moreover, solution (\ref{27}) can also represent regular black holes for $k\in (0,2M)$, but in this case WEC can be violated near $r=0$. Hence we have a
model that by changing the values of parameters
$k$ and $c$ can represent a particular interior source for Kerr metric or a regular black hole with a throat geometry at $r=0$. 
It is also interesting to note that the so obtained new class of solutions (\ref{2}) can 
also be used to find new black hole solutions sourronded by a dark halo, thus representing a possible model for a spiral galaxy.

\end{document}